# A Hierarchical Relationship between the Fluence Spectra and CME Kinematics in Large Solar Energetic Particle Events: A Radio Perspective

N Gopalswamy[1], P Mäkelä[1,2], S Yashiro[1,2], N Thakur[1,2], S Akiyama and H Xie[1,2]

[1]Code 671, NASA Goddard Space Flight Center, Greenbelt, MD 20771, USA
[2]The Catholic University of America, Washington DC 20064, USA

**Abstract.** We report on further evidence that solar energetic particles are organized by the kinematic properties of coronal mass ejections (CMEs)[1]. In particular, we focus on the starting frequency of type II bursts, which is related to the distance from the Sun where the radio emission starts. We find that the three groups of solar energetic particle (SEP) events known to have distinct values of CME initial acceleration, also have distinct average starting frequencies of the associated type II bursts. SEP events with ground level enhancement (GLE) have the highest starting frequency (107 MHz), while those associated with filament eruption (FE) in quiescent regions have the lowest starting frequency (22 MHz); regular SEP events have intermediate starting frequency (81 MHz). Taking the onset time of type II bursts as the time of shock formation, we determine the shock formation heights measured from the Sun center. We find that the shocks form on average closest to the Sun (1.51 Rs) in GLE events, farthest from the Sun in FE SEP events (5.38 Rs), and at intermediate distances in regular SEP events (1.72 Rs). Finally, we present the results of a case study of a CME with high initial acceleration (~3 km s$^{-2}$) and a type II radio burst with high starting frequency (~200 MHz) but associated with a minor SEP event. We find that the relation between the fluence spectral index and CME initial acceleration continues to hold even for this minor SEP event.

## 1. Introduction

The hierarchical relationship recently found between the fluence spectra of large solar energetic particle (SEP) events and the kinematics of the associated coronal mass ejections (CMEs) provides a strong evidence that energetic particles in these events are accelerated by CME-driven shocks [1], most likely by the diffusive shock acceleration process [2]. Shocks can form at any distance from the Sun depending on the speed of CMEs relative to the Alfven speed of the ambient medium. One of the earliest signatures of CME-driven shocks is type II radio bursts, which are due to the plasma emission process involving nonthermal electrons accelerated by the shocks [3]. The plasma emission occurs at the fundamental and second harmonic of the local plasma frequency in the vicinity of the shock, so the starting frequency of a type II burst indicative of the local plasma density and the distance from the Sun where such a density is prevalent. The starting frequency and the frequency range of the type II bursts thus provide important information on where the shocks form and how long they survive in the interplanetary (IP) medium [4]. The existence of a shock is a necessary condition for both type II radio bursts and SEP events because the same shock accelerates both ions and electrons, although the details of the acceleration process may differ for the two species [5]. The high starting frequencies of type II bursts are thus indicative of shock formation close to the Sun. The maximum energy of accelerated particles depends on the magnetic field strength, age of the shock, and the shock speed [2]. Near the Sun, the magnetic field strength is high, so particles can be accelerated to high energies. This is certainly the case in ground level enhancement (GLE) events, which are hard-spectrum events in which the shock forms at a heliocentric distance of ~1.5 Rs [6]. On the other hand, in large SEP events associated with CMEs involving filament eruption (FE) outside of active regions, the shock forms at large distances from the Sun as indicated by the lack of high-frequency (metric) type II bursts and the presence of IP type II bursts [7]. At large distances, the ambient magnetic field strength is low, so the acceleration efficiency proportional to the magnetic strength is also low, resulting in a lower maximum energy attained by the particles and hence a soft spectrum. A comparison of the shock formation heights between GLE and non-GLE events showed that shocks form closer to the Sun in the case of GLE events [8]. One can see that the high initial speed and high type II starting frequency imply high ambient magnetic field, which is a key factor in accelerating particles to high energies [9]. CMEs attain high speeds early on when the initial acceleration is high, become super-Alfvenic, and drive fast-mode shocks followed by the energization of electrons (type II radio bursts) and ions (SEP events). Such CMEs will have type II bursts with high starting frequencies. However, there are type II bursts starting at high frequencies, but lack high-energy SEP events. Accurate determination of shock formation heights became possible [10] during the early part of solar cycle 24 when data from the Solar Dynamics Observatory (SDO, [11]) became available along with data from the Solar Terrestrial Relations Observatory (STEREO, [12]). For a set of 32 CMEs, the shock formation height was found to be in the range 1.2 to 1.93 solar radii (Rs), with an average of 1.43 Rs [10]. Only 7 of the 32 CMEs had shock formation heights >1.5 Rs. None of the 32 CMEs was associated with a GLE event or a large SEP event. Only a handful of them were associated with minor SEP events with intensity <2 pfu (particle flux unit; 1 pfu = 1 particle per (cm$^2$ s sr)). The first event in [10] is the 2010 June 12 CME that had a >10 MeV intensity of <1 pfu.

One of the characteristics of the events similar to the 2010 June 12 eruption is the short duration of the associated soft X-ray flares and the rapid decline in CME speed similar to the X-ray time profile after attaining a high speed around the peak time of the soft X-ray flare [13]. Such events provide important constraints on the validity of the hierarchical relationship between the SEP spectrum and CME kinematics.

In this paper, we elaborate on the idea that the initial speed of a CME is a proxy to the high initial acceleration by describing the CME height-time history of a GLE event and an FE SEP event. We then illustrate that the grouping of SEP events into GLE, regular, and FE SEP events is also consistent with the type II starting frequencies and shock formation heights. Finally we study the CME kinematics, SEP association, type II burst association, and the fluence spectrum of an impulsively accelerated CME, which occurred on 2010 June 12 to illustrate that this event behaves like a regular SEP event.

## 2. CME Initial Speed and Acceleration

The reason for using the initial speed of the CMEs as a proxy is the lack of height-time measurements close to the Sun from the Large Angle and Spectrometric coronagraph (LASCO, [14]) on board the Solar and Heliospheric Observatory (SOHO, [15]) mission. The innermost coronagraph LASCO/C1 was capable of observing the acceleration profile of CMEs [16], but it operated for only less than 3 years during the rise phase of solar cycle 23. Most of the large SEP events occurred after LASCO/C1 ceased operations, so we have to rely on the outer coronagraphs C2 and C3. Here we illustrate how we can use the first two height-time data points from LASCO/C2 to assess the initial acceleration of CMEs. For this purpose, we use two events from cycle 24 that had both SOHO and STEREO observations to confirm the analysis.

*2.1. The 2012 May 17 GLE Event*
Figure 1 shows the height and speed variation of the 2012 May 17 CME, which produced the first GLE event in cycle 23 [17]. The CME first appeared in the LASCO/C2 field of view (FOV) at 01:48 UT at a height of 3.61 Rs. In the next frame taken ~12 minutes later, the CME leading edge was at 5.43 Rs, indicating a speed of ~ 1757 km s$^{-1}$. This must be close to the peak speed attained by the CME, but it started slowing down as is clear from the speed – time plot obtained by a second order fit to the height-time measurements. The last two data points are (24.27 Rs, 04:18 UT) and (25.80 Rs, 04:30 UT). The CME moved 1.53 Rs in 12 minutes corresponding to a speed of 1479 km s$^{-1}$. Clearly, the CME slowed down in the LASCO FOV as confirmed by the second order fit shown in Fig. 1b. Taking into account of the source location at N11W76, the deprojected initial and final speeds become 1810 and 1524 km s$^{-1}$, respectively. Fortunately, this event was observed by both SOHO and STEREO, so we were able to obtain the three-dimensional speed of the CME. The inner coronagraph COR1 and the Extreme Ultra Violet Imager (EUVI) in STEREO's Sun Earth Connection Coronal and Heliospheric Investigation (SECCHI, [18]) observed closer to the Sun providing the acceleration profile of the CME. Using multiview observations, it was found that the CME attained its maximum speed of 1977 km s$^{-1}$ at 02:00 UT and a peak acceleration of 1.77 km s$^{-2}$ at 01:36 UT. The initial speed from the first two LASCO/C2 data points is only slightly smaller than the peak speed obtained from the multiview observations. If we do not have CME observations near the Sun, we cannot measure the acceleration. However, we can get the average acceleration the CME undergoes from the initial CME speed (1810 km s$^{-1}$) and the flare rise time. The flare rise time in this event was 22 min, so we get an acceleration of 1.37 km s$^{-2}$, consistent with the peak acceleration. This simple analysis shows that the initial speed is indicative of the initial acceleration of CMEs.

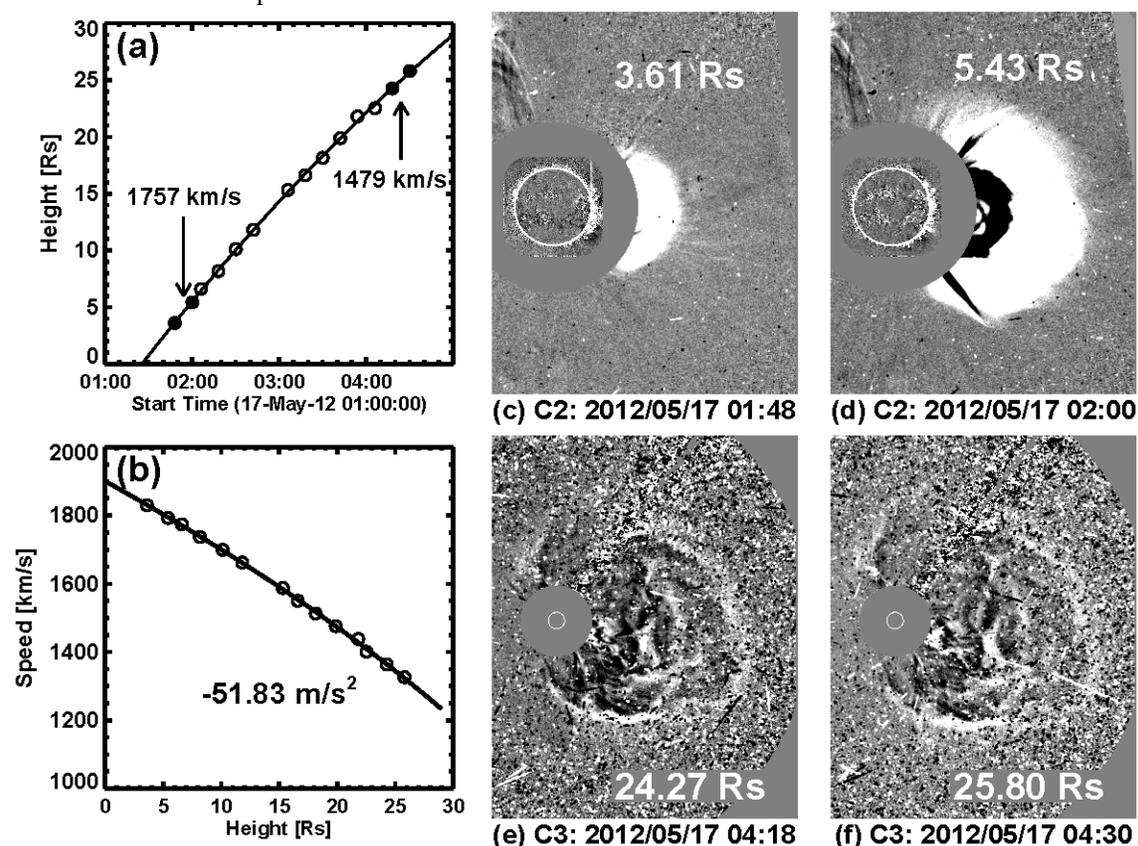

Figure 1. The height-time history of the 2012 May 17 GLE event (a) and the speed variation with height obtained using a second order fit to the height-time measurements (b). The speed obtained using the first two data points is ~1757 km s$^{-1}$, while the speed using the last two data points is ~1479 km s$^{-1}$, indicating a clear deceleration in the coronagraph FOV. The deceleration in (b) is the residual acceleration. LASCO difference images corresponding to the first two data points (c, d) and last two data points (e, f) are also shown. We have superposed difference images from SDO's Atmospheric Imaging Assembly (AIA, [19]) at 193 Å on LASCO/C2 images to indicate the eruption location on the Sun. The heliocentric distances of the CME leading edges are also given. The coronal images in (e) and (f) appear corrupted because of the energetic particles hitting the LASCO/C3 detectors.

The eruption was associated with a metric type II burst starting at 01:32 UT and the starting frequency was high (~200 MHz) for the harmonic component. Thus the plasma level at shock formation corresponds to a fundamental plasma frequency of ~100 MHz, indicating a shock formation close to the Sun (~1.38 Rs as derived from the CME height-time plot – see [10]). The CME speed was high enough to be super-Alfvenic and hence form a fast-mode shock. The solar particle release (SPR) was at 01:40 UT, about 8 minutes after the shock formation and the CME had reached a height of 2.32 Rs. Thus the shock travelled ~1 Rs after formation before particles were released.

*2.2. The 2011 November 26 FE SEP event*

Figure 2 shows the 2011 November 26 FE SEP event that had height-time and speed-time plots different from those of the GLE event in Fig. 1. The CME initial speed from the first two data points is 697 km s$^{-1}$, while the final speed in the LASCO FOV is 1402 km s$^{-1}$. The higher final speed is indicative of the continued acceleration of the CME. The centroid of the eruption region was N27W49, which can be used to get the deprojected initial and final speeds as 924 km s$^{-1}$ and 1858 km s$^{-1}$, respectively. A second order fit to all the height-time data points confirms the continued acceleration of the CME with an average acceleration of ~8.97 m s$^{-2}$. Observations from STEREO/COR1 showed that the initial acceleration of the CME was ~0.3 km s$^{-2}$. An extremely faint and narrow band type II burst started around 9 MHz in the Wind/WAVES [20] dynamic spectrum at 07:15 UT, corresponding to a shock formation height of ~3.5 Rs. The type II burst became intense only when the CME reached a height of ~13.5 Rs at 9:15 UT. Thus, the small initial speed is consistent with the low initial acceleration and accordingly the CME became super-Alfvenic only beyond the outer corona [7].

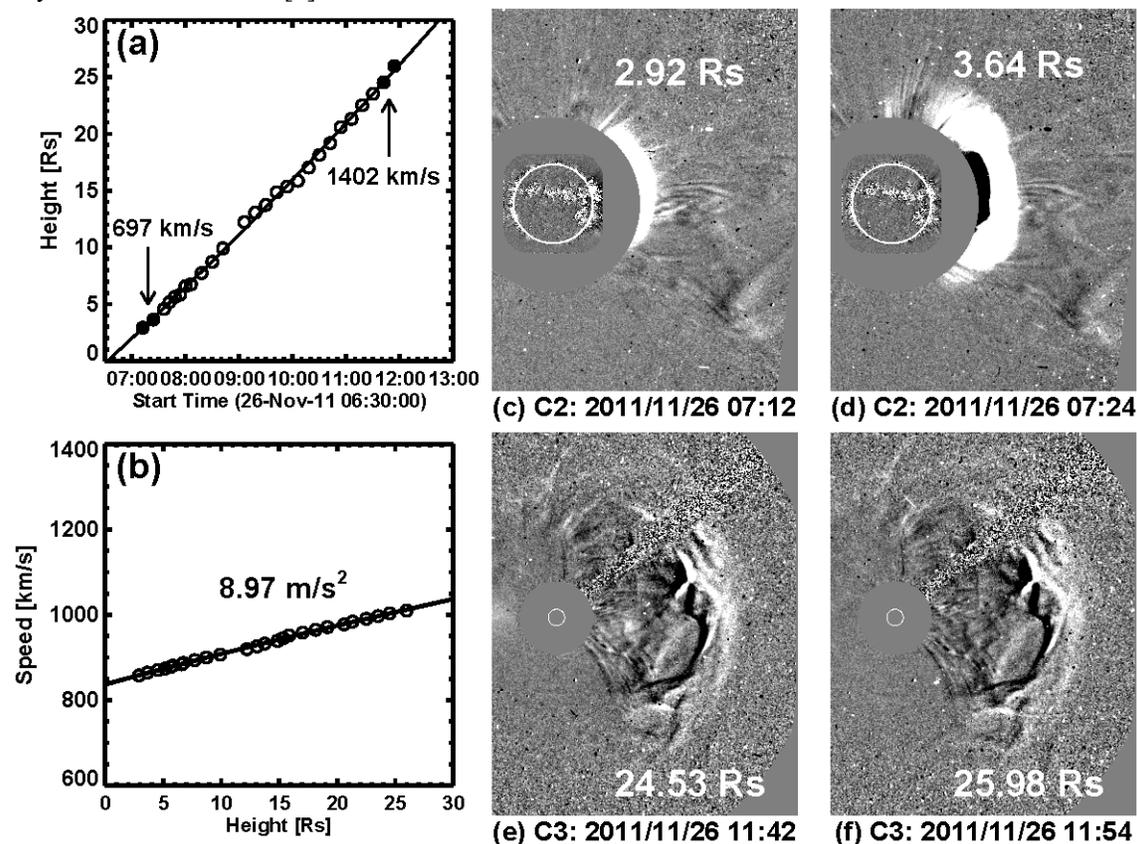

Figure 2. Same as Fig. 1, but for the FE SEP event of 2011 November 26. The initial speed using the first two data points is 697 km s$^{-1}$, while the speed from the last two data points is 1402 km s$^{-1}$, indicating that the CME continued to accelerate through the coronagraph FOV. This is confirmed by the positive acceleration in the LASCO FOV (8.97 m s$^{-2}$) obtained by a second order fit to the height-time measurements.

*2.3. Summary of comparison*

The above discussion provides a clear contrast between a GLE event and an FE SEP event, which differ in their initial acceleration, initial speed, and shock formation height (or starting frequency of the type II radio burst) as summarized in Table 1. Most parameters of the third category – the regular SEP events – fall between those of GLE and FE SEP events as was demonstrated in [1]. With this background we examine the starting frequencies of type II radio bursts and the derived shock formation heights for all SEP events of solar cycles 23 and 24 to understand the initial speed – initial acceleration relationship.

**Table 1.** Kinematics, shock formation height and fluence spectral indices of a GLE and an FE SEP events.

| Property | 2012 May 17 GLE | 2011 Nov 26 FE SEP |
|---|---|---|
| Initial speed (first 2 points) km $s^{-1}$ | 1757 | 697 |
| Final speed (last 2 points) km $s^{-1}$ | 1479 | 1402 |
| Average speed km $s^{-1}$ | 1582 | 933 |
| 3-D speed km $s^{-1}$ | 1997 | 1187 |
| Initial acceleration km $s^{-2}$ | 1.77 | 0.32 |
| Starting frequency of type II MHz | 100 | 9 |
| Shock formation height Rs | 1.38 | 3.5 |
| Fluence Spectral Index | 2.48 | 5.46 |

**3. Type II Radio Bursts Associated with Large SEP Events**

One of the main results of [1] is that the high initial CME speed is a good proxy to the initial acceleration of CMEs. High initial speed guarantees the formation of shocks near the solar surface. Since type II radio bursts are good indicators of CME-driven shocks in the corona, we can use the starting frequency of type II bursts as an indicator of the height at which the shocks formed.

We examined the dynamic spectra of type II bursts associated with large SEP events of cycle 23 and 24 and identified the type II starting frequencies. For determining the shock formation height, we need the starting frequency of the fundamental component. When both fundamental and harmonic components are observed in the dynamic spectrum, it is straight forward to obtain the starting frequency. If the fundamental component is present, but not well observed, we divided the harmonic frequency by 2 to get the starting frequency of the fundamental. If only a single component is present, we dropped the event. Another parameter we need is the onset time of the type II burst, which we take as the time of shock formation [10]. We then determined the leading edge of the CME at the onset time of the type II burst as the shock formation height. We assumed that the radio emission originates from the leading edge of the CME and hence ignore the possibility of radio emission coming from the shock flanks.

Figure 3a shows the distribution of the starting frequencies of type II radio bursts associated with the three types of large SEP events (GLE events, regular SEP events, and FE SEP events). Although there is significant overlap between the three groups of SEP events, we see that the GLE events have the highest average starting frequency, 107 MHz. Type II bursts associated with FE SEP events have the lowest starting frequency, 22 MHz, while those associated with regular SEP events have an intermediate starting frequency, 81 MHz. The shock formation height obtained by matching the onset time of the type II bursts with the CME leading edge height-time history is shown in Fig. 3b. Clearly the shock formation heights in GLE events are at the bottom of the plot (average 1.51 Rs). The shock formation heights of the regular SEP events overlap heavily with the GLE events and extend to larger heights. The average shock formation height is only slightly larger (1.72 Rs). The shock formation heights of the FE SEP events are generally large and have no overlap with those of GLE events. They also did not have overlap with the regular SEP events except for one event. The average shock formation height of FE SEP events is 5.38 Rs, much greater than those of the other two types.

According to the empirical relationship between the starting frequency of type II bursts and the heliocentric distance of the shock obtained in [10], the shocks should form at a heliocentric distance of 1.34 Rs for a starting frequency of 107 MHz in GLE events. Similarly, the starting frequencies of 81 MHz (regular SEP events) and 22 MHz (FE SEP events) correspond to heliocentric distances of 1.44 and 2.1 Rs, respectively. The actual measurements of CME leading edge heights are generally greater than these numbers as seen in Fig. 3b, although the hierarchical relationship is clearly preserved. It must be noted that the empirical relationship in [10] was derived from metric type II bursts that corresponded to shock heights <2 Rs. Here we have included type II bursts with starting at frequencies below 14 MHz. Even though the average starting frequencies of the three SEP populations are distinct, we see that there is considerable overlap between regular SEP events and GLE events. In particular, there are many high-starting frequency type II bursts in SEP events that are not GLEs. A preliminary look at these events shows that the shock speeds are not as high as that in typical GLE events. In some cases, the connectivity is poor either in latitude or longitude [21].

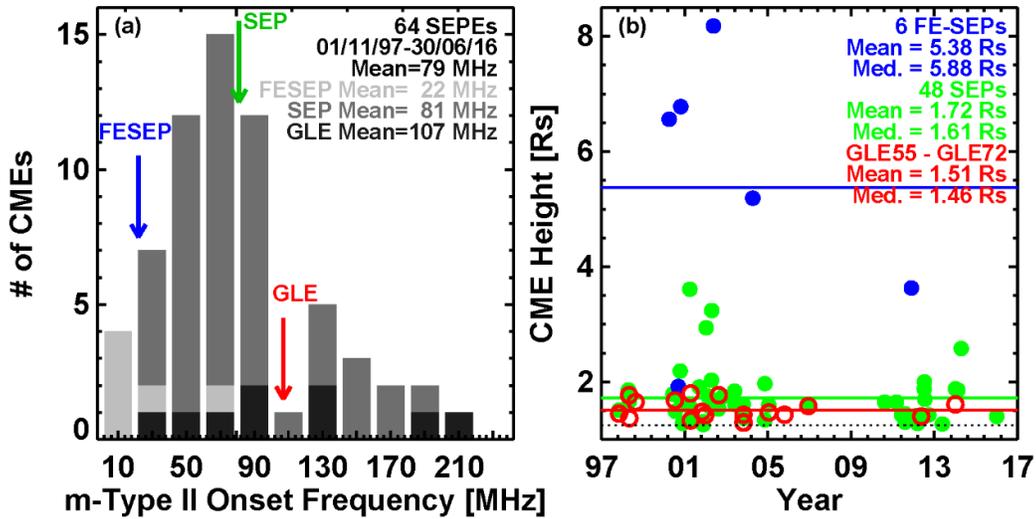

Figure 3. (a) Distribution of starting frequencies of type II bursts associated with large SEP events. The starting frequency is the lowest for FESEP events (average ~22 MHz), highest for GLE events (average ~107 MHz) and intermediate for regular SEP events (average ~81 MHz). The entire population of SEP events has a mean staring frequency of 79 MHz. (b) The shock formation height for the three types of SEP events: 5.38 Rs (FESEP events), 1.72 Rs (regular SEP events), and 1.51 Rs (GLE events). The shock formation height is consistent with the dispersion in starting frequencies among the three types of SEP events.

## 4. The 2010 June 12 Event

We consider the 2010 June 12 CME that was associated with a metric type II burst and a minor SEP event. The flare was of very short duration, lasting only for ~8 minutes between 00:54 and 01:02 UT. The type II burst started at 00:56 UT, just two minutes after the flare onset. Figure 4 shows the GOES soft X-ray flare profile and the type II radio burst. The type II burst has multiple lanes in the dynamic spectrum. We identify the highest frequency (~400 MHz) as harmonic emission. Therefore, the plasma level where the type II burst starts is at ~200 MHz. Previous estimate of the starting plasma frequency as 107 MHz was based on dynamic spectra with limited frequency extent [10]. Examining the extended dynamic spectrum from the Hiraiso Radio Spectrograph (HiRAS) [22], we found that the harmonic starts around 400 MHz, so we revised the fundamental plasma frequency to be ~200 MHz. The type III burst started at a frequency of ~2000 MHz, suggesting that the eruption occurred deep in the corona.

The CME associated with the eruption was observed as a slow but wide CME in the LASCO FOV. Following the analysis shown in Figs. 1 and 2, the first two LASCO data points give the initial speed is 476 km s$^{-1}$ and the last two data points give a speed of 412 km s$^{-1}$. These speeds are consistent with the average speed in the LASCO FOV (486 km s$^{-1}$) obtained by a linear fit to the height-time data points and an average deceleration of -5.2 m s$^{-2}$. Using the solar source location of this eruption (N22W57), we can deproject the speeds to 568 km s$^{-1}$ (initial) and 491 km s$^{-1}$ (final). The CME first appeared in the LASCO FOV at 01:31 UT, which is well after the end of the type II burst. This suggests that LASCO missed the initial phase of the CME that was driving a shock.

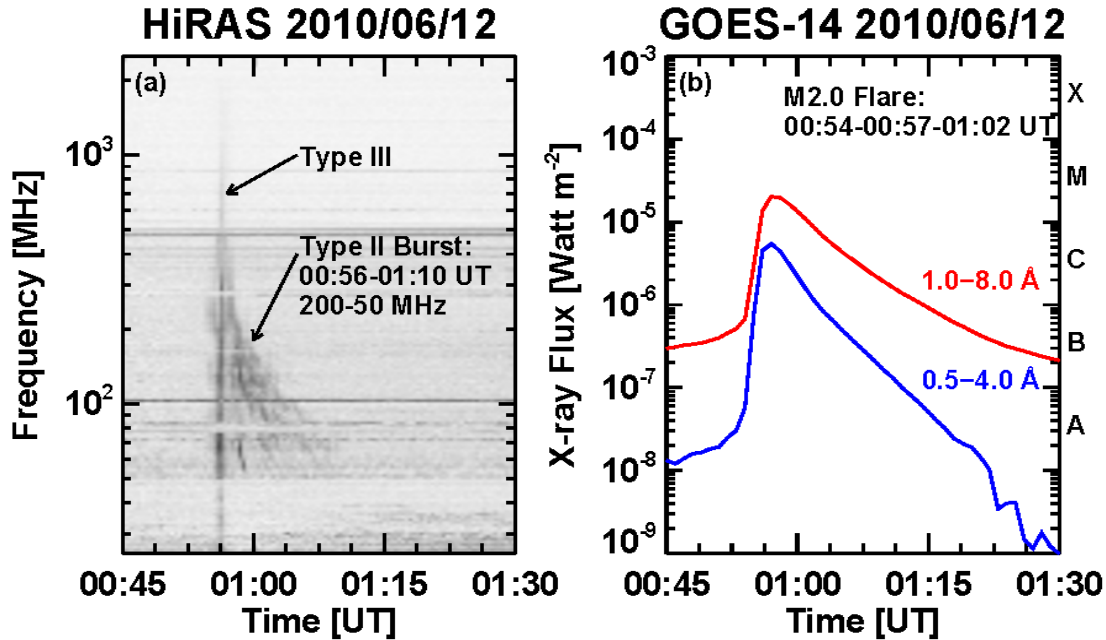

Figure 4. (a) Radio dynamic spectrum from the Hiraiso Radio Spectrograph (HiRAS) showing the type II and type III radio bursts. (b) The soft X-ray light curves of the M2.0 flare in two energy channels as observed by the GOES satellite.

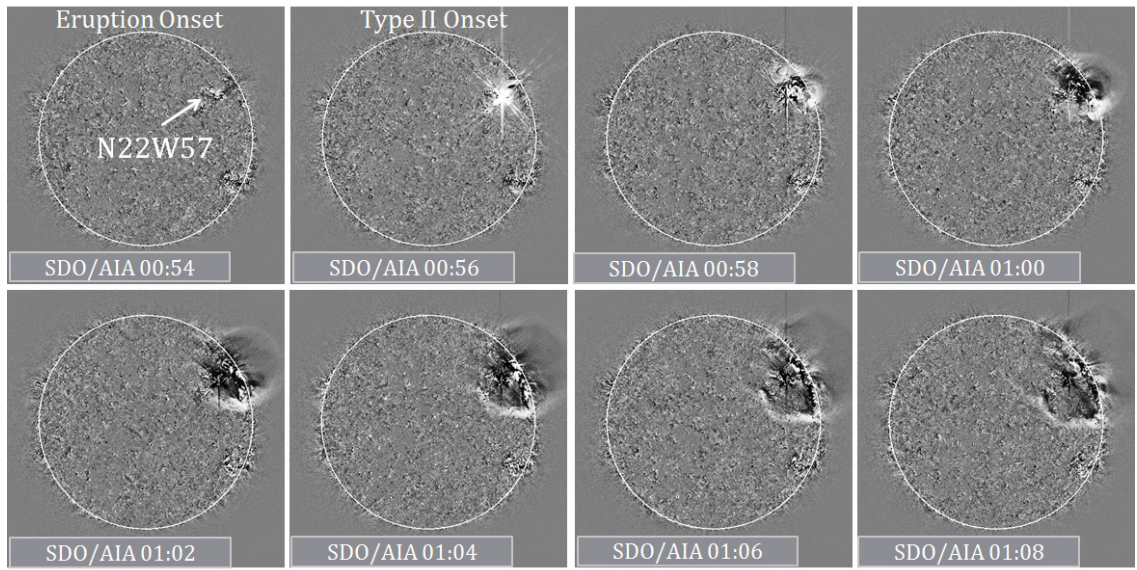

Figure 5. A series of SDO/AIA 193 Å difference images of the Sun on 2010 June 12 showing the EUV wave originating from the eruption site at N22W57. The snapshots are at every 2 minutes between 00:54 and 01:08 UT. The metric type II radio burst started when the leading edge of the wave was at the limb in projection, which corresponds to an actual height of 1.20 Rs from the Sun center. The type II burst ended around 01:10 UT.

Figure 5 shows the evolution of the eruption between 00:54 and 01:08 UT, roughly over the life time of the type II burst. A hint of the CME can be already seen in the 00:56 UT image extending to the limb in projection. Given the eruption location at N22W57, the CME leading edge can be estimated to be ~1.20 Rs, consistent with the high starting frequency. The empirical formula between the starting frequency ($f$ in MHz) of type II burst and the shock formation height ($r$ in Rs): $f = 308.17r^{-3.78} - 0.14$ [10], gives a shock formation height of ~1.13 Rs for $f = 200$ MHz in agreement with the measurements.

The bubble-like structure can be interpreted as the shock, although the flux rope is not seen distinctly. The shock was also observed as a well-defined circular front in STEREO-A EUVI close to the disk center (not shown). The CME was a limb event in STEREO-B FOV. Combining the SDO and STEREO-B/COR1 images, we were able to track the CME during its accelerating phase.

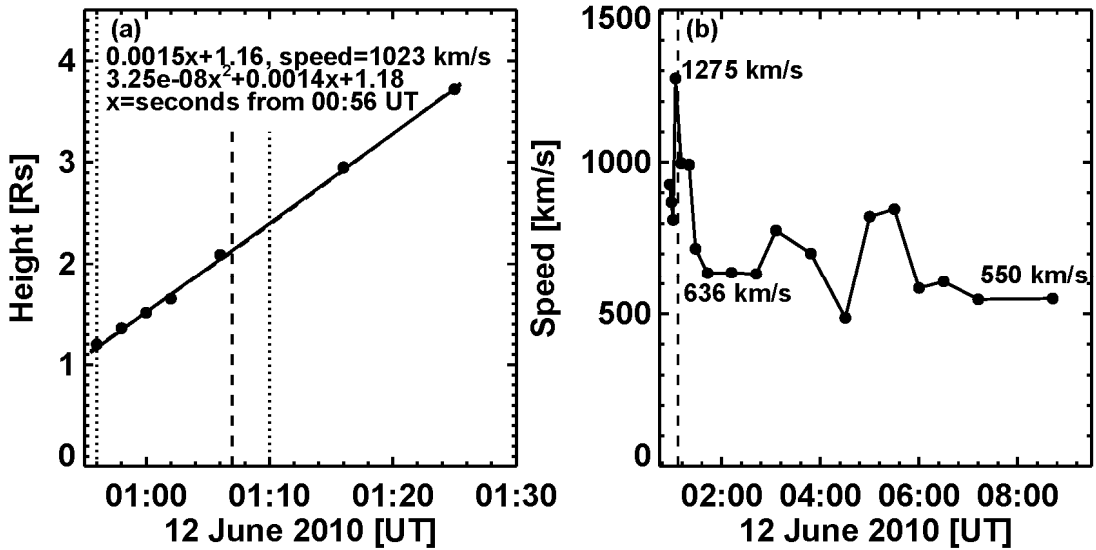

Figure 6. (a) The CME leading-edge height as a function of time obtained by combining SDO and STEREO/COR1 images. The solid vertical lines mark the onset and end of the type II burst. The dashed vertical line marks the time of solar particle release. (b) The CME leading-edge speed as a function of time obtained by combining SDO, STEREO, and SOHO images over an extended period of time. The vertical line at 01:07 UT marks the approximate solar particle release time, which is about 9 minutes after the onset of the type II burst.

*4.1. CME height-time history*
Figure 6 shows the CME height-time plot and the evolution of the leading edge speed. During the life time of the type II burst, the CME travelled ~1 Rs, from 1.2 to 2.2 Rs. The CME speed was obtained from successive height-time measurements and plotted as a function of time. When the type II burst started, the CME leading edge had already reached a speed of ~900 km s$^{-1}$ and further increased to a maximum value of ~1275 km s$^{-1}$. The speed dropped back to ~1000 km s$^{-1}$ when the type II ended. A similar event on 2007 June 3 with an impulsive soft X-ray flare and CME speed during the life time of the type II burst was reported in [13]. The main reason for the end of the type II in that event was an increase in the Alfven speed after the minimum around 1.5 Rs [23]. The evolution of the 2010 June 12 shock is also consistent with such a changing Alfven speed. The impulsive nature of the type II burst is indicative of the short-lived propelling force. The CME speed settled down around 600 km s$^{-1}$ through the SOHO/LASCO FOV and there was no IP type II burst. This suggests that either the shock died or it was too weak to accelerate significant number of electrons to produce type II radio emission.

*4.2. Energetic particles*

The 2010 June 12 eruption did produce an SEP event with a peak intensity of 0.3 particles per (cm$^2$ s sr MeV) in the 13.8 – 16.9 MeV energy channel of SOHO's Energetic and Relativistic Nuclei and Electron (ERNE) instrument [24]. The time profile of the SEP intensity and the fluence spectrum are shown in Fig. 7. The event lasted for about three days. The SEP event started at about 01:30 UT at Earth, which is about 34 minutes after the onset of the type II burst. Assuming a Parker spiral length of 1.2 AU, we estimated that 50 MeV protons would take ~31 minutes to reach Earth. Thus the particle release time at the Sun is 00:59 UT. Normalizing the travel time to the electromagnetic emissions (X-rays, type II radio emission), the solar particle release time becomes 01:07 UT, which is just a few minutes before the end of the type II burst (see Fig. 4). Thus the acceleration time for the 50 MeV protons is ~10 minutes, similar to other large SEP events.

The fluence spectrum of the 2010 June 12 SEP event is remarkably similar to those of regular SEP events [1]. The spectral index (3.14) lies between the 2001 April 12 regular SEP event and the 2004 April 11 FE SEP event. In fact, the spectral index of the 2010 June 12 SEP event is smaller than the average spectral index of all regular SEP events (3.83) reported in [1]. The plot between CME acceleration ($a$) and the fluence spectral index γ for large SEP events reported in [1] (their Fig. 8f) yields a regression line, γ = -0.29$a$ + 4.2. Substituting γ = 3.14, we get $a$ = 3.6 km s$^{-2}$. We can see that the initial acceleration of the CME is in this range because the CME speed increased from ~900 km s$^{-1}$ to 1275 km s$^{-1}$ in 2 min, which corresponds to $a$ = 3 km s$^{-2}$. Thus the 2010 June 12 SEP event, despite being a minor event, is consistent with the hierarchical relationship between CME acceleration and the fluence spectral index [1].

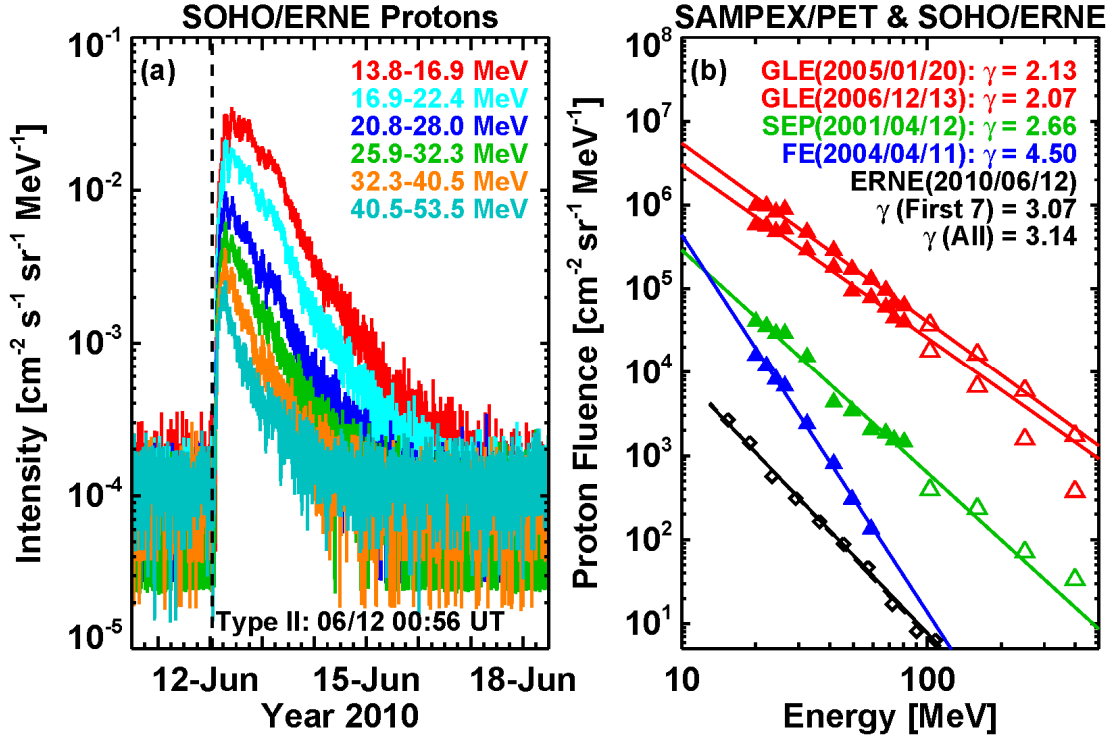

Figure 7. (a) Proton intensity of the 2010 June 12 event from SOHO/ERNE in several energy channels noted on the plot. The onset time of the type II burst is marked by the vertical dashed line. The onset time of the SEP event was around 1:30 UT. (b) Fluence spectrum of the 2010 June 12 event compared with a FE SEP, GLE, and regular SEP event from [1].

**5. Summary and Conclusions**

It has been recently found that CMEs with high initial acceleration are associated with SEP events with the hardest fluence spectra, while those with lowest initial acceleration have the softest SEP fluence spectra. Consistent with this trend, CMEs with intermediate initial acceleration have moderately hard SEP fluence spectra. Physically speaking, impulsive acceleration leading to high initial CME speeds close to the Sun results in shock formation close to the Sun, where the ambient magnetic field and density are high and the particles are energized more efficiently. On the other hand, slowly accelerating CMEs can form shocks only at several solar radii from the Sun, where the magnetic field and density have fallen off significantly, reducing the efficiency of shock acceleration. The GLE SEP events represent the high initial acceleration class, while the FE SEP events represent the low initial acceleration class. We illustrated this hierarchical relationship using the starting frequency of type II radio bursts associated with large SEP events in solar cycles 23 and 24. We found that the average starting frequency of type II bursts progressively decreases as one goes from GLE events (107 MHz), to regular SEP events (81 MHz) and to FE SEP events (22 MHz). Since the fundamental component of the type II burst is emitted at the local plasma frequency, and the local plasma frequency decreases with heliocentric distance, the frequency variation reflects the height of shock formation measured from the Sun center. We confirmed this by measuring the height of the CME leading edge at the time of type II burst onset assuming that the shock formation coincides with the onset of type II burst. We found that the shock formation distance increases as one goes from GLE events (1.51 Rs), to regular SEP events (1.6 Rs), and to FESEP events (5.38 Rs). This finding strongly supports the idea that large SEP events are primarily due to CME-driven shocks and reflect the changing ambient magnetic field strength influencing the efficiency of particle acceleration [9].

We also performed a case study of the 2010 June 12 CME, which had high initial acceleration, but the associated SEP event did not fall into the class of large SEP events because the intensity was <1 pfu. Interestingly, the relation between the initial CME speed and acceleration continued to hold. Despite the low SEP intensity, we found the fluence spectral index to be similar to that of regular SEP events. As in large SEP events, the energetic particles were released several minutes after the shock formation. However, the CME speed dropped abruptly, similar to the spiky soft X-ray profile, suggesting a short-lived propelling force. The sudden drop in the CME speed weakened the shock and the injection of the energetic particles into the heliosphere ended up being a delta function. A detailed analysis of such small SEP events and the associated type II bursts and CMEs will be reported elsewhere.

**Acknowledgments**
This work was supported by NASA's Heliophysics Guest Investigator program. PM was partially supported by NASA grant NNX15AB77G and NSF grant AGS-1358274. NT was partially supported by NSF grant AGS-1622377. HX was partially supported by NASA LWS TR&T grant NNX15AB70G.

## 6. References

[1] Gopalswamy N, Yashiro S, Thakur N, Mäkelä P, Xie H, and Akiyama S 2016 The 2012 July 23 Backside Eruption: An Extreme Energetic Particle Event? *Astrophys. J.* **833** 216

[2] Zank, G P, Rice, W K M., Wu, C C 2000 Particle acceleration and coronal mass ejection driven shocks: A theoretical model. *J. Geophys Res*. **105** 25079

[3] Melrose, D B 1987 Plasma emission - A review. *Solar Phys*. **111** 89

[4] Gopalswamy N, Xie H, Mäkelä P, Akiyama S, Yashiro S, Kaiser M L, Howard R A, Bougeret J-L 2010b Interplanetary Shocks Lacking Type II Radio Bursts. *Astrophys. J.* **710** 1111; Gopalswamy, N, Aguilar-Rodriguez, E, Yashiro, S, Nunes, S, Kaiser, M L Howard, R A 2005 Type II radio bursts and energetic solar eruptions. *J. Geophys Res*. **110**, A12S07

[5] Masters, A et al. 2013 Electron acceleration to relativistic energies at a strong quasi-parallel shock wave. *Nature Physics* **9** 164; Amano, T and Hoshino, M 2010 A critical Mach number for electron injection in collisionless shocks. *Phys. Rev. Lett*. **104** 181102; Lembege, B, Giacalone, J, Scholer, M, Hada, T, Hoshino, M Krasnoselskikh, V, Kucharek, H, Savoini, P, and Terasawa, T 2004 Selected Problems in Collisionless-Shock Physics. *Space Sci. Rev*. **110** 161

[6]     Cliver E W, Nitta NV, Thompson B J, Zhang J 2004 Coronal Shocks of November 1997 Revisited: The Cme Type II Timing Problem *Solar Phys*. **225** 105; Reames D V 2009 Solar Energetic-Particle Release Times in Historic Ground-Level Events *Astrophys. J.* **706** 844; Gopalswamy N, Xie H, Yashiro S, Akiyama S, Mäkelä P, Usoskin I G 2012 Properties of Ground Level Enhancement Events and the Associated Solar Eruptions During Solar Cycle 23. *Space Sci. Rev*. **171** 23; Desai, M and Giacalone, J 2016 Large gradual solar energetic particle events. *Living Rev. in Solar Phys* **13** 3

[7]     Gopalswamy N, Mäkelä P, Akiyama S, Yashiro S, Xie H, Thakur N, Kahler S W 2015 Large Solar Energetic Particle Events Associated with Filament Eruptions Outside of Active Regions *Astrophys. J.* **806** 8

[8]     Mäkelä P, Gopalswamy N, Akiyama S, Xie H, Yashiro S 2015 Estimating the Height of CMEs Associated with a Major SEP Event at the Onset of the Metric Type II Radio Burst during Solar Cycles 23 and 24 *Astrophys. J.* **806** 13

[9]     Zank, G P, Li, G, Florinski, V, Hu, Q, Lario, D, Smith, C W 2006 Particle acceleration at perpendicular shock waves: Model and observations. *J. Geophys Res*. **110** 6108; Li, G, Shalchi, A, Ao, X, Zank, G, Verkhoglyadova, O P 2012 Particle acceleration and transport at an oblique CME-driven shock. *Adv. Space Res*. **49** 1067

[10]    Gopalswamy et al. 2013 Height of shock formation in the solar corona inferred from observations of type II radio bursts and coronal mass ejections *Adv. Space Res*. **51** 1981

[11]    Schwer K, Lilly R B, Thompson B J, and Brewer D A 2002, The SDO Mission *AGU Fall Meeting Abstracts*, SH21C-01

[12]    Kaiser M L, Kucera T A, Davila J M St. Cyr O C Guhathakurta M, Christian E 2008 The STEREO Mission: An Introduction *Space Sci. Rev*. **136** 5

[13]    Gopalswamy N, Thompson W T, Davila J M, Kaiser M L, Yashiro S, Mäkelä P, Michalek G, Bougeret J-L, Howard R A 2009 Relation between type II bursts and CMEs inferred from STEREO observations. *Solar Phys*. **259** 227

[14]    Brueckner G E, Howard R A, Koomen M J, Korendyke C M, Michels D J, Moses J D 1995, The large angle spectroscopic coronagraph (LASCO), *Solar Phys*. **162** 357

[15]    Domingo V, Fleck B, and Poland A I 1995 SOHO: The Solar and Heliospheric Observatory *Space Sci. Rev*. **72** 81

[16]    Gopalswamy N and Thompson B J 2000 Early life of coronal mass ejections *JASTP* **62** 1457; Zhang J, Dere K P, Howard R A, Kundu M R, White S M 2001 On the Temporal Relationship between Coronal Mass Ejections and Flares *Astrophys. J.* **559** 452

[17]    Gopalswamy N, Xie H, Akiyama S, Mäkelä P, Yashiro S, Usoskin, I G, Davila, J M 2013 The first ground level enhancement event of solar cycle 24: direct observation of shock formation and particle release heights. *Astrophys. J.* **765** L30

[18]    Howard RA et al. 2008 Sun Earth Connection Coronal and Heliospheric Investigation (SECCHI). *Space Sci. Rev*. **136** 67

[19]    Lemen, J et al. 2012 The Atmospheric Imaging Assembly (AIA) on the Solar Dynamics Observatory (SDO) *Solar Phys*. **275** 17

[20]    Bougeret J-L et al. 1995 Waves: The Radio and Plasma Wave Investigation on the Wind Spacecraft *Space Sci. Rev*. **71** 231

[21]    Gopalswamy N, Xie H, Akiyama S, Mäkelä P, Yashiro S, Major solar eruptions and high-energy particle events during solar cycle 24 *Earth, Planets and Space* **66** 104

[22]    Kondo T, Isobe T, Igi S, Watari S, Tokumaru M 1995 The Hiraiso Radio Spectrograph (HiRAS) for Monitoring Solar Radio Bursts *J. Comm. Res. Lab*. **42** 111

[23]    Gopalswamy N, Lara A, Kaiser ML, Bougeret, J-L 2001 Near-Sun and near-Earth manifestations of solar eruptive events. *J. Geophys Res*. **106**, 25261

[24]    Torsti et al. 1995 Energetic Particle Experiment ERNE *Solar Phys*. **162** 505